\documentclass[aps,prl,twocolumn,showpacs,amsmath,amssymb,floatfix,superscriptaddress]{revtex4}
\usepackage{epsfig}
\usepackage{bm}

\begin{document}

\title{Equilibration of quantum Hall edge states by an Ohmic contact}

\author{Artur O. Slobodeniuk}
\affiliation{D\'epartement de Physique Th\'eorique, Universit\'e de Gen\`eve, CH-1211 Gen\`eve 4, Switzerland}
\affiliation{Bogolyubov Institute for Theoretical Physics,
             14-b Metrolohichna Street, Kiev 03680, Ukraine}

\author{Ivan P. Levkivskyi}
\affiliation{Department of Physics, Harvard University, Cambridge, MA 02138, USA}
\affiliation{Bogolyubov Institute for Theoretical Physics,
             14-b Metrolohichna Street, Kiev 03680, Ukraine}

\author{Eugene V. Sukhorukov}
\affiliation{D\'epartement de Physique Th\'eorique, Universit\'e de Gen\`eve, CH-1211 Gen\`eve 4, Switzerland}
\date{\today}

\begin{abstract}
Ohmic contacts are crucial elements of electron optics that have
not received a clear theoretical description yet. We propose a model of an Ohmic
contact as a piece of metal of the finite capacitance $C$ attached to a quantum Hall edge.
It is shown that charged quantum Hall edge states
may have weak coupling to neutral excitations in an Ohmic contact.
Consequently,  despite being a reservoir of neutral excitations,
an Ohmic contact is not able to efficiently equilibrate edge states
if its temperature
is smaller than $\hbar\Omega_c$, where $\Omega_c$ is the inverse RC time of
the contact. This energy scale for a floating contact may become as large as
the single-electron charging energy $e^2/ C$.
\end{abstract}

\pacs{72.70.+m, 73.23.-b, 73.22.-f, 42.50.Lc}
\maketitle

The study of quantum Hall (QH) edge states has
a great importance both for theoretical understanding of
a strongly correlated matter and
for the development of novel quantum devices for electron optics.
These chiral 1D states are
quantum analogs of the skipping orbits which appear at the edge
of a two dimensional electron gas (2DEG) in a strong magnetic field.
The fact that the QH edge states behave in many ways similarly to optical beams has triggered
several quantum optics-type experiments with electrons \cite{firstMZ,MZ2,MZ3,exper}.
One of the most important elements used to manipulate the edge states in such experiments
are the Ohmic contacts.
They serve as incoherent sources and detectors of the ``electron beams'' in experiments
on controllable dephasing \cite{MZ2,MZ3} and controllable energy equilibration of the edge states
\cite{exper}.

Ohmic contacts are created by placing a piece of metal on top of
a highly doped region in a semiconductor containing 2DEG. Strong tunneling between the edge
states and the states in this doped region provides a low resistance contact with external circuits.
Ohmic contacts are very complicated electron systems from the theoretical point of view, and they are still the least
understood elements of electron optics in spite of their widespread usage.
A simplification arises in the regime of integer QH effect where
edge states are commonly described using the free fermion picture \cite{fermion}.
Based on this picture,
an idealized concept of a so-called ``voltage probe'' \cite{but}
has been proposed to describe floating Ohmic contacts in this regime.
Voltage probe is a reservoir of electrons, which absorbs all incoming electron excitations
and emits new electron states with the equilibrium Fermi distribution and
the electrochemical potential that takes into account the current conservation law.
\begin{figure}[]
\includegraphics[width=8cm]{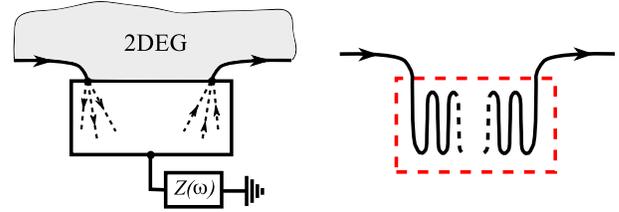}
\caption{An Ohmic contact, shown on the left, is a piece of metal (white rectangle)
which is placed in close proximity to the 2DEG (shown by gray shadow) and
connected to an electrical circuit via the impedance $Z(\omega)$. It absorbs
incident electron states, formed at the edge of the 2DEG in the QH effect regime
(shown by thick black lines) and turns them to neutral electron-hole excitations
(dotted lines). Then, it emits equilibrium neutral excitations, and at the exit, turns
them to the edge states.  Motivated by earlier findings \cite{chamon,averin}, 
we model this process by extending an edge state
inside the piece of metal (red rectangle shown on the right), and splitting it in two uncorrelated
channels. Edge states inside the metal provide their charge $Q$ to the Ohmic contact and become neutral modes.
The Ohmic contact, in turn, equilibrates these neutral modes.}
\vspace{-5mm}
\label{Ohmic0}
\end{figure}
However, it has been shown recently \cite{Sukh-Che,Chalker,ourPRB,others1,others2}, in the context of the experiments
\cite{MZ2,exper},  that even at integer filling factors the free-fermion description of the edge states
is not always correct,
and that the effective theory \cite{wen} considering the edge states as collective boson excitations
is a more appropriate approach.
This observation calls to
reconsider the applicability of the concept of a voltage probe.

It may appear that this concept finds a theoretical justification  even
at the effective theory level.
Indeed, it
has been argued that strong tunnel coupling at a floating Ohmic contact leads to equilibration
\cite{chamon} and an  effective
elongation \cite{averin} of the QH edge channel. Therefore it seems to be natural
to consider the edge channel
being effectively cut into two separate parts  (see Fig.\ \ref{Ohmic0}) carrying orthogonal fermions,
i.e., with zero overlap. However, edge states carry
the electric charge, which an Ohmic contact may have a limited ability to accommodate.
To be more specific, considering an Ohmic contact as a three-dimensional piece of metal of the size $L$,
the level spacing of neutral excitations scales as  $1/L^3$. It is typically small enough to consider
an Ohmic contact to be a reservoir of such excitations. On the other hand, the characteristic frequency
of the charge response of an Ohmic contact scales as $1/C$, where its capacitance $C$ is of the order of $L$. Therefore,
it scales down with the size $L$ much slower than the level spacing and may
compare to characteristic energy scales of modern mesoscopic experiments with QH edge states, which
makes it impossible to fully equilibrate them.
In this Letter,  we propose a simple model of an Ohmic contact,
 generalizing earlier models for systems with  non-chiral Luttinger liquids \cite{earliermod},
which is capable to correctly account its finite charge response frequency.

\begin{figure}[]
\includegraphics[width=4.5cm]{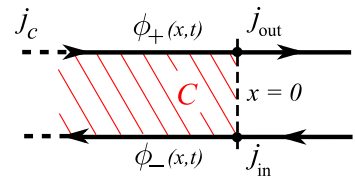}
\caption{An equivalent representation of the floating Ohmic contact at filling factor $\nu = 1$
shown in Fig.\ \ref{Ohmic0}. For convenience, we fold edge states
so that they could be described by two boson fields $\phi_+$ and $\phi_-$
of opposite chiralities. The dynamics of these fields is generated by the Hamiltonian (\ref{ham-all}), and
the boundary conditions
are given by Eqs.\ (\ref{BC}). The region inside the Ohmic contact, where the capacitive
interaction is assumed, is shown by the red color. Note that the charge response frequency of the contact
is finite, while the level spacing of neutral modes vanishes. In order to take this
fact into account, we extend edge states inside the interaction region to infinity and introduce a small
parameter $\epsilon$ to regularize corresponding integrals.}
\vspace{-5mm}
\label{Ohmic1}
\end{figure}

{\it Floating contact and boson scattering theory.}---We
model an Ohmic contact connecting incoming and outgoing QH edge states at filling factor $\nu=1$
as shown in Fig.\ \ref{Ohmic1} and explained in the figure caption.
The low-energy physics of the QH edge states \cite{wen}
is described by a set of scalar boson fields $\phi_{\sigma}(x,t)$,
where $\sigma=\pm$. The charge
density operator and the current operator for incoming, $\sigma = -$, and outgoing, $\sigma = +$, states
have the form $\rho_{\sigma}=({e}/{2\pi})\partial_x\phi_{\sigma}$, and $j_{\sigma}=-({e}/{2\pi})\partial_t\phi_{\sigma}$.
These boson fields satisfy the following canonical commutation relations:
\begin{equation}
[\partial_x\phi_{\sigma}(x,t),\phi_{\sigma'}(y,t)]=2\pi i\sigma \delta_{\sigma\sigma'}\delta(x-y).
\label{comm}
\end{equation}
The system can be described by the Hamiltonian containing two parts.
The first part generates the dynamics of the incoming and outgoing edge channels \cite{footnote0},
while the second term
describes the charging energy of the Ohmic contact of a finite size:
\begin{subequations}
\label{ham-all}
\begin{eqnarray}
\mathcal{H} &=& \frac{\hbar v_F}{4\pi}\sum_{\sigma}\int_{-\infty}^\infty \!\!\! dx (\partial_x\phi_{\sigma})^2
+\frac{Q^2}{2C}, \label{ham} \\
Q &=& \int_{-\infty}^0 dx e^{\epsilon x/v_F}[\rho_{+}(x)+\rho_{-}(x)].\label{qu}
\end{eqnarray}
\end{subequations}
Here $Q$ is an operator of the total charge accumulated at the Ohmic contact, and
$\epsilon$ is a small regularization parameter, roughly given by the decay rate of neutral excitations
inside the contact \cite{model}. 

Using commutation relations (\ref{comm}) and the Hamiltonian (\ref{ham-all}),
we obtain the equations of motion for the incoming and outgoing state:
\begin{equation}
\sigma\partial_t\phi_\sigma(x,t)+ v_F\partial_x\phi_\sigma(x,t)= -\frac{e}{\hbar C}Q(t)e^{\epsilon x/v_F}\theta(-x).
\label{eom}
\end{equation}
These equations have to be accompanied with the boundary conditions
\begin{subequations}
\label{BC}
\begin{eqnarray} \label{BC-L}
\partial_t\phi_+(-\infty,t) &=& -(2\pi/e){j}_{c}(t),\\
\partial_t\phi_-(0,t) &=& (2\pi/e)j_{\rm in}(t),\label{BC0}
\end{eqnarray}
\end{subequations}
where  $j_{\rm in}$ is the current flowing into the Ohmic contact, while ${j}_{c}$ describes {\em equilibrium}
fluctuations of the neutral mode with the reservoir temperature $T_c$, originating from the Ohmic contact.
Solving the equations (\ref{eom}) with the boundary conditions (\ref{BC}), one can
relate the outgoing current $j_{\rm out}(t)=-(e/2\pi)\partial_t\phi_+(0,t)$ to the incoming current $j_{\rm in}$,
as illustrated in Fig.\ \ref{Ohmic1}.

In order to solve Eqs.\ (\ref{eom}), we apply the Fourier transform
$\phi_\sigma(x,\omega)\equiv\int dt e^{i\omega t}\phi_\sigma(x,t)$ and
rewrite them as ordinary first order differential equations.
The general solution for $x\leq 0$ then reads
\begin{equation}
\phi_\sigma(x,\omega)\!=\!
\phi_\sigma(\omega)e^{i\sigma\omega x/v_F}+ \frac{\sigma}{R_qC}
\frac{\sum_{\sigma'}\phi_{\sigma'}(\omega)}{(i\omega- \sigma\epsilon)}e^{\epsilon x/v_F},
\label{solution}
\end{equation}
where $\phi_\sigma(\omega)$ are constants of integration, and $R_q=2\pi\hbar/e^2$ is the
resistance quantum.
Using the boundary conditions (\ref{BC}), we find the fields $\phi_\sigma(x,t)$, and   then,
the outgoing current
\begin{equation}
\label{floating_current}
j_{\rm out}(\omega)=\!\frac{i\omega R_qC}{i\omega R_qC-1}{j}_{c}(\omega)-
\frac{1}{i\omega R_qC-1}j_{\rm in}(\omega),\!
\end{equation}
where we have omitted a trivial phase factor in $j_c(\omega)$
and set $\epsilon=0$ \cite{footnote1}.
In the context of the boson scattering theory \cite{Sukh-Che,scattering},
the boundary conditions (\ref{BC}) can be viewed as the incident waves,
while $j_{\rm out}$ is the outgoing wave.
Then the coefficients in front of the currents in Eq.\ (\ref{floating_current})
are the boson scattering amplitudes \cite{footnote2}.

{\it Langevin equations.}---The equations of motion for the currents and the charge may be written
in a yet different form:
\begin{subequations}
\begin{eqnarray}
\frac{dQ(t)}{dt}&=&j_{\rm in}(t)-j_{\rm out}(t),
\label{cons1}\\
j_{\rm out}(t)&=&Q(t)/R_qC+j_{c}(t),
\label{total_current}
\end{eqnarray}
\label{LE0}
\end{subequations}
where the first equation expresses the conservation of charge. The second one is
the Langevin equation, which has the following simple physical meaning. The outgoing
current acquires two contributions: $Q(t)/R_qC$ is the current induced
by the time dependent potential $Q(t)/C$, and the second one, $j_c$, is the Langevin
current source. It is easy to check that by solving these equations, one arrives at the result
(\ref{floating_current}).

\begin{figure}[tb]
\includegraphics[width=5cm]{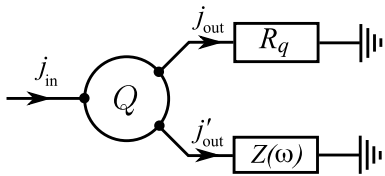}
\caption{Equivalent circuit representation of the Langevin equations (\ref{LE}). The charge conservation
in the system is described by Eq. (\ref{LE-cons}) , while Eqs.\ (\ref{LE-cur1}) and (\ref{LE-cur2})
for the outgoing currents
are the Langevin equations with the current sources $j_c$ and $j_Z$, respectively.}
\vspace{-5mm}
\label{Ohmic4}
\end{figure}

The advantage of this formulation is that the equations (\ref{LE0}) can be easily generalized
to account for the effects of dissipation in an Ohmic contact connected
to electric circuit.
This amounts to
adding a current $j'_{\rm out}$ and an impedance $Z$ to the equivalent
electric circuit, as shown in Fig.\ \ref{Ohmic4}. It is convenient to present corresponding equations
of motion in the frequency domain:
\begin{subequations}
\label{LE}
\begin{eqnarray}
i\omega Q(\omega) &=& j_{\rm out}(\omega)+j'_{\rm out}(\omega)-j_{\rm in}(\omega),\label{LE-cons}\\
j_{\rm out}(\omega) &=& Q(\omega)/R_qC+j_c(\omega),\label{LE-cur1}\\
j'_{\rm out}(\omega) &=& Q(\omega)/Z(\omega)C+ j_Z(\omega).\label{LE-cur2}
\end{eqnarray}
\end{subequations}
After straightforward calculations, we present the current $j_{\rm out}(\omega)$ in the following form
\begin{subequations}
\label{current}
\begin{eqnarray}
j_{\rm out}(\omega) \!&= &\!\sum\nolimits_{p = {\rm in}, c, Z}\mathcal{T}_p(\omega)j_p(\omega), \\
\mathcal{T}_Z = -\mathcal{T}_{\rm in}\!&=&\! \mathcal{T}_{c} - 1\nonumber\\
\!&=&\![i\omega R_qC-R_q/Z(\omega)-1]^{-1}.
\label{Ts}
\end{eqnarray}
\end{subequations}
We note, that the used here Langevin equation approach is appropriate in the case of linear quantum circuits that 
may be described entirely in terms of collective plasmon modes. This approach is fully consistent  \cite{Ingold:book} 
with the Caldeira-Leggett model \cite{CL}.

{\it Spectral functions and effective temperature.}---
One can characterize the statistics of the current fluctuations $\delta j_{\rm out}\equiv j_{\rm out} - \langle j_{\rm out}\rangle$
by the spectral density function, $S(\omega)$, defined via the relation
\begin{equation}
\langle \delta j_{\rm out}(\omega)\delta j_{\rm out}(\omega') \rangle=2\pi\delta(\omega+\omega')S(\omega).
\end{equation}
The solution (\ref{current}) allows us to express it
in terms of the noise spectral densities of the
currents $j_{\rm in}(\omega)$, $j_c(\omega)$, and $j_Z(\omega)$:
\begin{equation}
\label{spectral_density_function}
S(\omega)=\sum_p |\mathcal{T}_p(\omega)|^2S_p(\omega).
\end{equation}
Here $S_p(\omega)$
are defined similarly, $\langle \delta j_p(\omega)\delta j_p(\omega') \rangle=2\pi\delta(\omega+\omega')S_p(\omega)$,
and the average is evaluated with the equilibrium state in the corresponding channel, which implies  \cite{Landau:book9}
\begin{equation}
\label{equilibrium_spectral_densities}
S_p(\omega)=\frac{2\hbar \omega G_p}{1-e^{-\hbar\omega/T_p}},
\end{equation}
with $G_{\rm in}=G_c=1/2R_q$, and $G_Z={\rm Re}[1/Z(\omega)]$.
It is easy to check that the following identity holds:
\begin{equation}
\sum_p |\mathcal{T}_p(\omega)|^2G_p(\omega)=1/2 R_q.
\label{srule}
\end{equation}
Therefore, the statistics of the current originating at the Ohmic contact is equilibrium
if all the temperatures $T_p$ are equal, and it is {\em non-equilibrium}
otherwise.

\begin{figure}[]
\includegraphics[width=8cm]{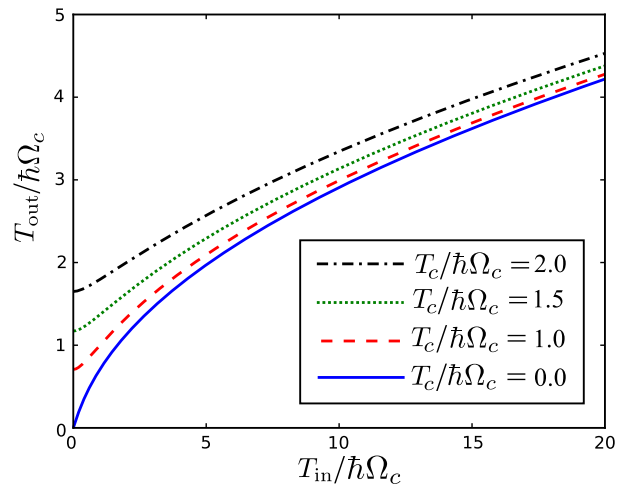}
\caption{The effective temperature $T_{\rm out}$ of the QH edge state at $\nu=1$ originating at
a floating Ohmic contact is plotted as a function of the temperature $T_{\rm in}$ of the incoming edge state
for different values of the temperature of the Ohmic contact $T_c$.
Note, that at $T_{\rm in}=0$ the effective temperature of the outgoing channel saturates at a constant value,
which is smaller than the contact's temperature $T_c$. At large values of $T_{\rm in}$, it behaves as
$T_{\rm out}\propto\sqrt{T_{\rm in}}$. }
\vspace{-5mm}
\label{Ohmic2}
\end{figure}

Sufficiently far from the Ohmic contact the outgoing state reaches the equilibrium.
Therefore, since the heat flux $J_{\rm out}$ carried by the edge state is conserved, 
it is convenient to define the effective
temperature $T_{\rm out}$ of the outgoing state by using the equilibrium relation
$J_{\rm out}=\pi T_{\rm out}^2/12\hbar$ for a 1D chiral channel \cite{footnote3},
which can be easily derived as follows \cite{our-rel}. 
In a chiral system, the energy flux operator is equal to the energy density operator multiplied by the velocity
$J=v_F\cdot ({\hbar v_F}/{4\pi})[\partial_x\phi(x,t)]^2=({\pi\hbar}/{e^2})j^2(x,t)$. 
Then, the heat flux can be obtained by subtracting the vacuum energy contribution: 
$J_{\rm out}=\langle J\rangle-J_{\rm vac}$.
Expressing the heat flux in terms of the current noise spectral function and comparing it to the one for 
the equilibrium noise,  we arrive at the following result
\begin{equation}
\label{effective_temperatures}
T^2_{\rm out}=\frac{3R_q\hbar}{\pi^2}\int_{-\infty}^\infty
\!\!d\omega [S(\omega)-\hbar\omega\theta(\omega)/R_q].
\end{equation}
One can see, that the summation rule (\ref{srule}) guarantees the convergence of the integral in (\ref{effective_temperatures})
at high frequencies. 

We evaluate this integral using Eqs.\ (\ref{Ts}), (\ref{spectral_density_function}), and
(\ref{equilibrium_spectral_densities}) in the simple case where the circuit is a resistor, $Z(\omega)=R$, and
with the natural assumption $T_Z=T_c$. The result reads
\begin{equation}
T^2_{\rm out}=T_c^2+\frac{6(\hbar\Omega_c)^2}{(\pi\gamma)^2}\Big[ I\Big(\frac{\hbar\Omega_c}{T_{\rm in}}\Big)-
I\Big(\frac{\hbar\Omega_c}{T_c}\Big)\Big],
\label{main1}
\end{equation}
where $\Omega_c=(R_q+R)/R_qRC$ is the inverse RC time of the Ohmic contact, and
$\gamma=1+R_q/R$ is the circuit coupling parameter.
The dimensionless function $I$ in this equation has the following form
$$
\label{integral}
I(a)=\!\!\int_0^\infty\!\!\!\frac{zdz}{z^2+a^2}
\frac{1}{e^z-1}=\frac12\left[\ln\left(\frac{a}{2\pi}\right)-\frac{\pi}{a}
-\psi\left(\frac{a}{2\pi}\right)\right],
$$
where $\psi(z)$ is the logarithmic derivative of the Gamma function.
The Fig.\ \ref{Ohmic2} shows $T_{\rm out}$ for the floating contact as
a function of $T_{\rm in}$ for different values of $T_c$.

In the case of a cold Ohmic contact, $T_c=0$, we find
\begin{equation}
\frac{T_{\rm out}}{T_{\rm in}}=\left\{
\begin{array}{ll}
1/\gamma, &\mbox{if $T_{\rm in}\ll \hbar \Omega_c$},\\
\sqrt{3 \hbar\Omega_c/\pi\gamma^2 T_{\rm in}}, &\mbox{if $T_{\rm in}\gg \hbar\Omega_c$}.
\end{array}
\right.
\end{equation}
Note, that for $\gamma =1+R_q/R>1$ additional cooling is provided by the dissipation in the circuit.
In the case of a cold incoming state, $T_{\rm in}=0$, and finite $T_c$, we have
\begin{equation}
\frac{T_{\rm out}}{T_c}=\left\{
\begin{array}{ll}
\sqrt{1-1/\gamma^2}, &\mbox{if $T_c\ll  \hbar\Omega_c$},\\
1-3  \hbar\Omega_c/2\pi\gamma^2 T_c, &\mbox{if $T_c\gg  \hbar\Omega_c$}.
\end{array}
\right.
\end{equation}
Thus, the ability of the Ohmic contact to equilibrate the edge state depends on the energy scale $\hbar\Omega_c$,
which, for a floating contact ($\gamma=1$), becomes comparable to the single-electron charging energy: 
$\hbar\Omega_c=e^2/2\pi C$. To efficiently equilibrate edge states with temperatures, e.g.,  in
the range $T_{\rm in} \sim 10-100$~mK,
one needs an Ohmic contact of the size of $L\sim 10-100$~$\mu$m or larger. 

So far, the contact's temperature $T_c$ has been considered an independent parameter, taking (ideally) 
the value of the base temperature $T_b$. The ability of phonons to cool a contact to the base temperature  
can be estimated by comparing the incoming heat flux of electrons $\pi T_{\rm in}^2/12\hbar$ to the 
outgoing flux $\Sigma V(T^5_c-T^5_b)$ to phonons, where $\Sigma \simeq 0.2$ nW/$\mu$m$^3$/K$^5$ 
is the electron-phonon coupling constant in metals \cite{phonons}, and $V\simeq L^3$ is the volume
of the contact. Assuming, that cooling by phonons 
is efficient, we find the relative correction $(T_c-T_b)/T_c\simeq \pi/60\hbar\Sigma(T_cL)^3$,
where we took $T_{\rm in}=T_c$ for the estimate. On the other hand,  
$\hbar\Omega =e^2/2\pi C\simeq e^2/2\pi\varepsilon\varepsilon_0 L$, where 
$\varepsilon\simeq 12$ for GaAs. Eliminating $L$, we find that  
$(T_c-T_b)/T_b\simeq (0.03\hbar\Omega_c/T_c)^3$, i.e., in the regime $T_c\simeq \hbar\Omega_c$ 
considered here heating of the Ohmic contact is indeed weak.  

{\it Multichannel case.}---The generalization of our model to QH systems with the integer filling factors
$\nu>1$ is straightforward. In this case, the interactions at the edge split the spectrum of the collective modes
in one charged mode and $\nu-1$ neutral modes.
For a floating contact, one may generalize the scattering theory. However, the easiest way to proceed is
by noting that one should simply replace the resistance $R_q$ in the Langevin equation (\ref{LE-cur1})
with $R_q/\nu$. Next, all the outgoing neutral modes are,  obviously, at equilibrium with the Ohmic contact,
because only the charged mode is coupled
to its charge $Q$. Finally, even a screened Coulomb interaction at the edge is typically strong enough
to equally distribute the heat flux over the $\nu$ electron channels \cite{our-rel}
that are accessible experimentally. All this leads to the modification of the circuit parameter
$\gamma=1+R_q/\nu R$, to a similar change in the charge response frequency
$\Omega_c=(R_q+\nu R)/R_qRC$, and to the overall suppression of the heat flux per channel.
Expressed in terms of the new parameters, the effective temperature of an electron channel is given by
$T^2_{\rm out}=T_c^2+(6/\nu)(\hbar\Omega_c/\pi\gamma)^2[ I(\hbar\Omega_c/T_{\rm in})-
I(\hbar\Omega_c/T_c)]$. In particular, in this case even at small temperatures $T_{\rm in},T_c\ll \hbar\Omega_c$
a floating Ohmic contact is able to heat the edge states ($T_{\rm out}/T_c=\sqrt{1-1/\nu}$ at $T_{\rm in}=0$)
or cool them down  ($T_{\rm out}/T_{\rm in}=1/\sqrt{\nu}$ at $T_c=0$) by redistributing the energy
uniformly over the electron channels.

To conclude, we have shown that a floating Ohmic contact attached to the edge of a QH system can serve
as a voltage probe only if it has a sufficiently large capacitance, so that the energy scale $\hbar\Omega_c$,
where $\Omega_c$ is the inverse RC time of the contact, is much
smaller than the temperature of the QH edge excitations. Such Ohmic contacts can be used to cool an edge
channel with temperature larger than $\hbar\Omega_c$, and the efficiency of cooling can be increased
by connecting the contact to a circuit with small resistance
$R < R_q$. Finally, the equilibration by an Ohmic contact becomes more efficient at large filling factors.

We acknowledge support from the Swiss NSF.


\begin{thebibliography}{99}
\bibitem{footnote0}
It is convenient
and safe to assume, for simplicity, that outside the Ohmic contact the Coulomb potential
is screened on a distance shorter than the characteristic experimental length scales, and
therefore, it merely increases the Fermi velocity $v_F$.
\bibitem{model}
In a similar model, proposed earlier by Flensberg and Matveev \cite{Flensberg},
a quantum dot is connected to a quantum point contact,  
which is replaced by a fictitious 1D system of free fermions.
\bibitem{footnote1}
We take the limits in Eqs.\ (\ref{BC-L}) and (\ref{solution}) as
$x\to-\infty$, and then $\epsilon\to 0$. Therefore,
the second term in Eq.\ (\ref{solution}) vanishes, so that $j_c$ is the current of
the neutral mode.
\bibitem{footnote3}
We note, that the commonly used definition of the effective temperature is based on the Nyquist formula $T=R_qS(0)$. 
However, for the floating Ohmic contact the so defined temperature is always equal to $T_{\rm in}$. 
Therefore, we use the alternative definition (\ref{effective_temperatures}).
\end{thebibliography}
\end{document}